\documentclass[
preprint,
runinaddress,
preprintnumbers,
nofootinbib,
 amsmath,amssymb,amsthm,
 aps,
prd,
]{revtex4-1}

\usepackage{lipsum}
\usepackage{dcolumn}
\usepackage{bm}
\usepackage{tikz}
\usetikzlibrary{matrix}
\usepackage{slashed}
\usepackage{fontenc}
\usepackage{graphicx,hyperref,color,xcolor}
\usepackage[scr=rsfso,cal=zapfc,frak=euler,bb=ams]{mathalfa}

\definecolor{awesome}{rgb}{1.0, 0.13, 0.32}
\definecolor{electricblue}{rgb}{0.03, 0.57, 0.82}
\definecolor{guppiegreen}{rgb}{0.0, 0.88, 0.4}
\definecolor{blue-violet}{rgb}{0.54, 0.17, 0.89}

\makeatletter
\def\p@figure{\color{awesome}}
\def\p@equation{\color{electricblue}}
\makeatother

\hypersetup{
    bookmarks=true,         
    unicode=false,          
    pdftoolbar=true,        
    pdfmenubar=true,        
    pdffitwindow=false,     
    pdfstartview={FitH},    
    pdftitle={Breakdown of quantum mechanics in gravitational holography},    
    pdfauthor={A. Akal},     
    pdfnewwindow=true,      
    colorlinks=true,       
    linkcolor=awesome,          
    citecolor=blue-violet,        
    filecolor=blue,      
    urlcolor=blue-violet          
}

\usepackage{titlesec}
\titleformat*{\section}{\large}
\titleformat*{\subsection}{\itshape}
\titleformat*{\paragraph}{\bfseries\itshape}

\begin{document}

\preprint{YITP-22-74}
\title{\textmd{\Large{Breakdown of quantum mechanics in gravitational holography}}}
\author{I. Akal}
\email[Current e-mail: ]{a.akal@uu.nl}
\thanks{Current affiliation: Institute for Theoretical Physics, Utrecht University, Princetonplein 5, 3584 CC Utrecht, The Netherlands}
\affiliation{Center for Gravitational Physics and Quantum Information\\
Yukawa Institute for Theoretical Physics, Kyoto University\\
Sakyo-ku, Kyoto 606-8502, Japan\\}
\date{\today}

\begin{abstract}
According to the holographic principle, the information content assigned to a gravitational region is processed by its lower dimensional boundary. As an example setup compatible with this principle, the AdS/CFT correspondence relies on the existence of D-branes in superstring theory. Black hole complementarity is inevitably linked to holography and states that information associated with the collapsed pure state is reflected in the near horizon region. Yet, if this is so, it is indispensable to understand the mechanism that makes black holes viewed from the outside evolve unitarily. We here argue that the information preserving quantum atmosphere of the black hole emerges from hidden variables on its horizon which would necessitate going beyond a probabilistic description within standard quantum theory. In AdS/CFT, this would mean that the completion of the semiclassical subalgebra to the complete boundary algebra has to be traced back to the emergent near horizon Hilbert space structure. The present investigations suggest that spacetime horizons, in general, may play a crucial role in restoring a long speculated ontology in quantum mechanics.
\end{abstract}

\maketitle

\section{Motivation}
\label{sec:mot}

What is the logic underlying the laws of nature processing information? At macroscopic and subatomic scales, the universe is described with an enormous precision by general relativity and quantum mechanics, respectively. While gravity is entirely deterministic, i.e.~classical, and nonlinear in its formulation, quantum theory is nonclassical, linear, and intrinsically probabilistic, at least according to its standard, but not incontestable Copenhagen interpretation. 

Do we, therefore, have to understand the world separately in classical and quantum logical terms? Where, admittedly, it is not really clear what would be meant by logic in the latter case. If yes, this sounds odd and clashy. If not, we do not know what the underlying rules look like, but they should give rise to today's theoretical machinery capable of describing the world to a certain level.

Formulating the physics of black holes requires both general relativity and quantum mechanics. A rather straightforward approach, however, results in a tension with the principle of unitarity \cite{Hawking:1974rv, Hawking:1975vcx}. It has been speculated whether unitarity is a valid concept in the presence of gravity \cite{Hawking:1976ra}.\footnote{Refer, for instance, also to \cite{Penrose:1996cv}.} 

A way out of this shortcoming is provided by the holographic principle \cite{Bekenstein:1972tm,Bekenstein:1973ur,tHooft:1993dmi, Susskind:1994vu}. Given a gravitational region, it argues that the associated information is processed by its lower dimensional boundary. In other words, the entropy assigned to that region should scale with its boundary surface area, not its volume.

As a model compatible with the holographic principle, the AdS/CFT correspondence \cite{Ma, Gubser:1998bc, Witten:1998qj} is realized via D-branes in superstring theory according to which gravitational physics in anti-de Sitter (AdS) spacetime is dual to a lower dimensional conformal field theory (CFT) on the asymptotically flat AdS boundary. 
In its stronger form, type IIB string theory on AdS$_5$ $\times$ $S^5$ is conjectured to be dual to $\mathcal{N}=4$ supersymmetric Yang--Mills theory on flat four dimensional spacetime. 

D-branes also play a substantial role in the microscopic derivation \cite{Strominger:1996sh} of the Bekenstein--Hawking entropy formula $S_\text{BH} = A_h/ (4 \ell_\text{P}^2)$, where $\ell_\text{P} = \sqrt{G \hbar/c^3}$ is the Planck length and $A_h = 4 \pi r^2_\text{S}$ denotes the black hole's horizon area being proportional to the Schwarzschild radius $r_\text{S}$ squared \cite{Bekenstein:1973ur, Hawking:1975vcx}. 

The idea of black hole complementarity \cite{tHooft:1984kcu, tHooft:1990fkf, Susskind:1993if} is closely linked to the holographic principle. In fact, it must be implied by the latter if, as expected, local quantum field theory is valid in the physical exterior region. Given a black hole formed from a pure state by collapse, complementarity is the statement that all the initial information is reflected in the near horizon region.\footnote{Notably, here, we discard the original statements involving the interior of the black hole.\label{fnote:comp}} 

In an infalling observer scenario, complementarity is argued to be safe due to the chaotic, scrambling near horizon dynamics \cite{Hayden:2007cs,Sekino:2008he}. In the context of string theory, the chaotic nature of the near horizon region \cite{Susskind:1993if} may give rise to the notion of a stretched horizon \cite{Price1986MembraneVO}, thereby realizing the idea of a thermal black hole atmosphere \cite{Zurek:1985gd, tHooft:1984kcu, tHooft:1996rdg}. Configurations of strings within the stretched horizon may account for the entropy of the black hole \cite{Susskind:1994sm}. On the other hand, the direct derivation of the Bekenstein--Hawking formula in three dimensional AdS spacetime \cite{Strominger:1997eq} is based on earlier findings pointing out that the asymptotic symmetry group of AdS$_3$ is generated by two copies of the Virasoro algebra \cite{Brown:1986nw}.\footnote{In effect, many of the recent studies on the relationship between geometry and quantum information in the context of holography do not exploit more than this insight.}

An excited, single long string state can be interpreted as the quantum dual of the black hole system \cite{Susskind:1993ws,Horowitz:1996nw}. Information may escape via the decay of the long string. If black holes viewed from the outside behave like ordinary quantum systems, as suggested by the holographic principle and string theory, their radiation entropy has to evolve in accordance with unitarity \cite{Page:1993df, Page:1993wv}. 

A unitary entropy curve in AdS/CFT turns out to require additional information in the near horizon region that cannot be captured semiclassically \cite{Akal:bhinter}. Such near horizon correlations, making the quantumness of the black hole seem blown up (in Planckian units), may determine the entropy evolution in general \cite{Akal:qes}. 

Importantly, reshuffling quantum information to the microscopic vicinity of the horizon evades the usual shortcomings of a global, semiclassical background. In fact, as it is stated in the present article, information may only reemerge in the exterior due to peculiar properties of the black hole horizon, see also \cite{Akal:prep}. This implies that the required information will not be existent in the interior, cf.~e.g.~\cite{tHooft:1999imj,tHooft:1999xul}. Notably, according to the entropic understanding of gravity \cite{Jacobson:1995ab,Padmanabhan:2009vy,Verlinde:2010hp}, some information loss ought to be in place. 

We here particularly argue for the compatibility of a nonprobabilistic, hidden variable description explaining the proposed emergence of the near horizon quantum mechanics. We further elaborate that the corresponding hidden variables are operative on or even across the sub-Planckian vicinity of the black hole horizon. The present explorations suggest that spacetime horizons, in general, may play a significant role in formulating a long speculated ontology in quantum mechanics.

The remainder of the paper is organized as follows. 

In Sec.~\ref{sec:inter}, we comment on the black hole interior in AdS/CFT. In Sec.~\ref{sec:qm}, we discuss general aspects of nonprobabilistic approaches to quantum mechanics. In Sec.~\ref{sec:tangle-hvs}, we propose the notion of emergent hidden variable independence and present arguments in favor of a hidden variable formulation in gravitational holography. In Sec.~\ref{sec:exwigner}, we elaborate on a recent thought experiment and its implications. In Sec.~\ref{sec:eqbh}, we further comment on the necessity of going beyond standard quantum theory and provide additional motivations for why the proposed hidden variables need to be operative on the black hole horizon, thereby supporting the arguments put forward in Sec.~\ref{sec:tangle-hvs}. We conclude in Sec.~\ref{sec:conc}.

\section{Black hole interior in AdS/CFT}
\label{sec:inter}

Since an infalling observer should pass through the horizon, we resign from considerations which deviate from a no-drama scenario. The dynamics of a quantum black hole viewed from outside is described within AdS/CFT, so that the near horizon aspects discussed before will apply. A convenient way of formulating the underlying physics is the language of operator algebras. Indeed, in addressing black hole unitarity in the holographic context, the algebraic perspective suggests the importance of (non-semiclassical) correlations across the quantum atmosphere \cite{Akal:bhinter}. 

In what follows, we comment on the black hole interior in AdS/CFT. For present purposes, we do not delve into the details to keep the discussion brief. Let $\mathcal{F}$ be an algebra, arguably a von Neumann algebra, which we refer to as the complete CFT algebra. When we say complete, we mean that $\mathcal{F}$ preserves bulk-boundary unitarity according to the holographic dictionary. The Hilbert space $\mathcal{H}_\mathcal{F}$ associated with $\mathcal F$ is viewed to be nonfactorized. 

The bulk consists of the black hole system surrounded by the physical spacetime. Any quantum theory is seen as a collection of algebras. Whether unitarity is preserved or not depends on the operator content (i.e.~representation) at hand. We stick to the Heisenberg picture. The state may correspond to the global wave function. Given a state, different choices for the algebra may result in distinct Hilbert spaces having differing spacetime support.

We may consider a subalgebra of $\mathcal{F}$. For instance, take a set of operators $O$ that belong to the algebra $\mathcal{N} \subset \mathcal{F}$ for which $\langle E_i | O | E_j \rangle \simeq \exp \left( -S(E) \right)$, where $i \neq j$. The states $E_i$ correspond to microstates defined in the boundary theory, where $S(E)$ denotes the average entropy depending on the central charge of the CFT. One may further degrade $\mathcal{N}$ to some closed subalgebra $\mathcal{N}'$ with elements $O'$ satisfying $\langle E_i | O' | E_j \rangle = 0$. We call any operation under the action of $\mathcal{N}'$ semiclassical. 

Under such semiclassical operations, the pure boundary state $| \Psi \rangle =  \sum_i  \alpha_i | E_i \rangle$ reduces to a thermal density matrix due to the existence of superselection sectors. The latter is dual to the AdS black hole according to the standard dictionary, thus, showing how, globally, the interior of the black hole may emerge from reducing the operator algebra \cite{Akal:bhinter}. We should note that simple, low complexity operators in AdS/CFT acting in the exterior would not strictly close under a subalgebra, instead, they generally belong to an algebra of the sort $\mathcal{N}$. This should be understood as a sign of complementarity.\footnote{Here, in the modified sense discussed in Sec.~\ref{sec:mot}, see footnote \ref{fnote:comp}.}

Consider now the case of two noninteracting CFTs being maximally entangled with each other forming a thermofield double (TFD) state \cite{Israel:1976ur}. In the bulk, the TFD state is viewed to be dual to a two sided eternal AdS black hole spacetime with its left and right boundaries described by CFT$_L$ and CFT$_R$, respectively \cite{Maldacena:2001kr}. 

However, as mentioned above, maximally entangling the boundary CFTs means maximally correlating their complete algebras $\mathcal{F}_L$ and $\mathcal{F}_R$. Taking into account the arguments above, their correlations should accordingly not rely on the black hole interior. Quantum information associated with the two systems, instead, has to be reflected in the near horizon regions on both sides. 

One may of course get back to the eternal AdS black hole with its semiclassical interior, as one requires from the infalling observer's perspective. But the cost for this would be reducing $\mathcal{F}_{L}$ and $\mathcal{F}_{R}$ to the respective subalgebras. This, however, will imply that unitarity cannot be maintained. In other words, due to the restricted algebra accessible to the infalling observer, information will be lost. 

The scenario sketched above deviates from the picture in \cite{Maldacena:2013xja} that we find misleading in light of the prerequisite of matching between the underlying operator algebras. In particular, we do not adopt the assertion EPR $\Rightarrow$ ER. Although further arguments have appeared in favor of the latter, see~e.g.~\cite{Susskind:2014yaa}, the adduced situation, however, closely relates to the procedure based on the holographic entropy proposal in AdS/CFT \cite{RT}. 

What happens is that one reduces $\mathcal{F}$ in order to arrive at a semiclassical bulk spacetime that is dual to the boundary state under the action of $\mathcal{N}$. In such a case, the holographic prescription can be followed to compute the amount of quantum correlations which, however, must physically rely on $\mathcal{F}$ in order to guarantee unitarity and, thus, prevent information loss. Under the action of the latter, the notion of a global spacetime would surely cease to exist. 

Eventually, from the outside perspective, the interior has to be regarded as nonphysical but may serve as a computational tool. However, this requires careful treatment. One must necessarily bear in mind, that being able to perform certain computations may not automatically imply the physical justification of the resulting picture.
  
\section{Quantum mechanics may be complete, but ...}
\label{sec:qm}

One may reasonably feel uncomfortable with the interpretational completion of quantum mechanics. Nonetheless, among various approaches, the Copenhagen interpretation has been the most widely accepted one. It may be reassuring if one repeatedly stumbles across the counterintuitive aspects of the quantum world. The Copenhagen interpretation basically asserts the following. 

(\textbf{I}) Quantum mechanics is intrinsically probabilistic, i.e. nondeterministic.

(\textbf{II}) Quantum mechanics leads to classical physics in the appropriate limit. This is known as the correspondence principle.

(\textbf{III}) The wave function $\psi$ in quantum mechanics yields probabilities $p$ for measurement outcomes according to the Born rule, i.e.~$p = |\psi|^2$.
This can be rephrased by a positive operator valued measure (POVM) acting on a finite dimensional Hilbert space, so that $p_i = \mathrm{tr} (\rho O_i) = \langle \psi | O_i | \psi \rangle$
yields the probability $p_i$ for obtaining the measurement outcome $i$ associated with the POVM element $O_i$, i.e.~a positive semi definite matrix, with respect to the pure quantum state $\rho = | \psi \rangle \langle \psi |$.

(\textbf{IV}) Definitions of certain properties at the same time are excluded. This has been crucial in Bohr's response to the Einstein--Podolsky--Rosen (EPR) argument \cite{EPR}.   

\subsection{Quantum measurement}

The wave function $\psi$ in quantum mechanics evolves according to the linear Schr{\"o}dinger equation. The measurement upon the system described by $\psi$ may change the state of the system. One says that the measurement has collapsed the wave function. For instance, one may decompose an POVM element $O_i = K_i^\dag K_i$, where $K_i$ are called Kraus operators specifying the measurement process. The Kraus operators may not be self adjoint, but their combinations $K_i^\dag K_i$ are. After the measurement, the initial state $\rho$ would reduce to
$\rho' = \frac{K_i \rho K_i^\dag}{\mathrm{tr}(\rho O_i)}$. In the case when the POVM is a projection valued measure (PVM), the Kraus operators act as projection operators $\Pi_i = \Pi_i^2$.
The state after the projective measurement becomes $\rho' = | i \rangle \langle i |$, when the initial state $\rho$ is pure and the projectors $\Pi_i = | i \rangle \langle i |$ have rank one. 

Now, consider a measurement process in which the measurement apparatus is brought into an eigenstate $| a \rangle$ with probability $p=1$, whereas the prepared initial state shall be $| s \rangle$, such that $\rho_s \rightarrow \rho_a$. Similarly, for another initial state $\rho_{s'}$, the apparatus is brought into the eigenstate $\rho_{a'}$ with $p=1$ after the measurement. Take an initially superimposed state, say $| \psi \rangle = \alpha | s \rangle + \beta | s' \rangle$. After performing the measurement in a given basis, the apparatus, which itself should be described by quantum mechanics, will be brought into one specific eigenstate with certain probability. 

More specifically, with the measurement basis chosen as $\{ |s\rangle, |s'\rangle \}$ in the present example, the probability would be either $|\alpha|^2$ or $|\beta|^2$. The apparatus will not end up being in a superposition state. In other words, the nonlinearity of the measurement process conflicts with the linearity of the Schr{\"o}dinger equation. This is known as the measurement problem. We here do not worry about the irreversibility of the measurement. In fact, this is rather practical and not a principle hurdle.

\subsection{Hidden variables}

Numerous interpretational attempts have been made to overcome the controversies related to the intrinsic probabilistic formulation of quantum mechanics. A particular example is provided by deterministic approaches known as hidden variable theories, which introduce hypothetical entities that may circumvent the indeterminacy constrained by uncertainty and uncover the reality that underlies quantum mechanics. These attempts may be divided into local and nonlocal hidden variable theories. 

The de Broglie--Bohm theory, also known as Bohmian mechanics, is a prominent nonlocal hidden variable theory \cite{Bohmian}; a so-called pilot wave theory. In addition to the probabilistic wave function, which evolves according to the Schr{\"o}dinger equation as in conventional quantum theory, one also introduces an existent configuration space for the particles whose configurations evolve in time according to a guiding equation. 

In the latter case, there will be no measurement problem since the configuration space is considered to be factual. The configuration space may vary depending on the type of the pilot wave theory. For instance, its elements can be thought of as the position variables for the particles. The statistical distribution of the particles may be determined according to the Born rule, which would define a state of quantum equilibrium. 

On the other hand, based on the usual assumptions, any deterministic hidden variable theory consistent with quantum mechanical predictions has necessarily to be nonlocal. This follows from Bell's theorem \cite{Bell:theorem} and stands at odds with the locality principle. Hidden variable theories, therefore, are considered to be ruled out by experimentally verified violations of the Bell inequalities due to quantum entanglement. However, under certain modifications of the original starting point, violations of the inequalities may not exclude a local, deterministic description \cite{Bell2004-BELSAU, Brans1988}. So far, a proposal realizing this scenario, however, has not been much convincing.

\subsection{Bell's theorem}

Let $\lambda$ be a continuous, factual, i.e.~not quantum mechanical, variable. It may even represent a set of (discrete) variables or include even more complications. Consider measurements made on selected components of spins $\sigma_1$ and $\sigma_2$. 

For instance, $A(a | \lambda)$ shall denote the measurement outcome effected by the parameter $\lambda$ of measuring $\sigma_1 \cdot a$, where $a$ denotes the apparatus setting. Similarly, the value $B(b | \lambda)$ represents the outcome of measuring $\sigma_2 \cdot b$ determined by the apparatus setting $b$ and the parameter $\lambda$. In each apparatus the measured particle selects one of two channels labeled by $+1$ and $-1$, so that $A(a|\lambda) = \pm 1$ and $B(b|\lambda)=\pm 1$. 

The experimenters may be located arbitrarily far from each other. As locality is assumed, one has apparatus setting independence, i.e.
\begin{equation}
A(a,b | \lambda) = A(a |\lambda)\quad
\wedge\quad
B(a,b | \lambda) = B(b | \lambda),
\label{eq:asi}
\end{equation}
means, that measurement outcomes in one apparatus do not depend on the setting of the other apparatus. Furthermore, the normalized probability distribution of the hidden variable(s), $\rho$, is assumed to be independent of the apparatus settings $a$ and $b$, hence
\begin{equation}
\rho(\lambda | a,b) = \rho(\lambda).
\label{eq:hvi}
\end{equation}
We call this hidden variable independence. 

Bell's theorem \cite{Bell:theorem} asserts that under the assumptions in Eqs.~\eqref{eq:asi} and \eqref{eq:hvi}, the quantum mechanical expectation value $\langle \sigma_1 \cdot a\ \sigma_2 \cdot b \rangle = - a \cdot b$
for the singlet state cannot be of the form
\begin{equation}
P(a,b) = \int d\lambda\ \rho(\lambda) P(a,b|\lambda),
\label{eq:P-hvt}
\end{equation}
where 
\begin{equation}
P(a,b|\lambda) = A(a | \lambda) B(b | \lambda).
\label{eq:fact}
\end{equation}
Based on such assumptions, the mismatch with quantum mechanics can also be expressed in terms of the Clauser--Horne--Shimony--Holt (CHSH) inequality \cite{CHSH},
\begin{equation}
P(a,b) + P(a',b) + P(a',b') - P(a,b') \leq 2,
\label{eq:chsh}
\end{equation}
where the experimenters may choose between apparatus settings $a,a'$ and $b,b'$, respectively.
Quantum mechanics violates the CHSH inequality in Eq.~\eqref{eq:chsh} by means of an upper bound $2 \sqrt{2}$ known as Tsirelson bound \cite{Cirelson}.

In quantum mechanics, violations of the Bell inequalities occur as the assumption of measurement outcome independence, means, that the distribution of $A$ does not depend on the distribution of $B$, i.e.
\begin{equation}
P(a,b | \lambda) = A(a,b | \lambda) B(a,b | \lambda),
\label{eq:moi}
\end{equation}
is not respected. Taking the formal relation in Eq.~\eqref{eq:moi}, expressing measurement outcome independence, together with Eq.~\eqref{eq:asi}, yields the factorization condition in Eq.~\eqref{eq:fact}. The latter is assumed in Eq.~\eqref{eq:P-hvt}, which is sometimes referred to as Bell separability.

The nonlocal feature of any factual theory required to recover the predictions of quantum mechanics can be related to the notion of quantum contextuality through the Bell--Kochen--Specker (BKS) theorem \cite{Bell2, KS}, assuming that the Hilbert space dimension is greater than two. 
The BKS theorem states that any noncontextual hidden variable theory cannot reproduce the predictions of quantum mechanics.

\subsection{A loophole}

Alike the Bell and BKS no-go theorems, the Pusey--Barrett--Rudolph (PBR) theorem \cite{PBR} rules out realistic, objective attempts explaining the predictions of quantum theory. For instance, take two distinct pure quantum states $|\Psi \rangle$ and $|\Psi' \rangle $. Let $\rho$ and $\rho'$ be the corresponding distribution functions describing the probability of the states being in some physical state represented by $\lambda$. 

The PBR theorem says that if $\rho$ and $\rho'$ have an overlap, that is, if the wave function is Psi-epistemic (state of knowledge about reality) \cite{harrigan2010einstein} as opposed to being Psi-ontic (state of reality), then there appears a contradiction with quantum mechanics. More precisely, assuming that the states represent mere information (knowledge) about the real physical state, two independently prepared states would not be in a product state. They have to be correlated, which would violate preparation independence. 

On the other hand, various loopholes have been identified in the Bell inequalities \cite{larsson2014loopholes}. Relevant to the present discussion are the principally unavoidable cases. A local, hidden variable theory may bypass the assumptions underlying Bell's theorem and thus violate the CHSH inequality if one of the assumptions in Eqs.~\eqref{eq:asi} and \eqref{eq:hvi} does not hold. Remember that Eq.~\eqref{eq:moi} is violated by quantum mechanics.

A possibility that cannot be put aside is respecting apparatus setting independence (i.e.~locality condition), Eq.~\eqref{eq:asi}, but violating hidden variable independence, Eq.~\eqref{eq:hvi}, i.e.
\begin{align}
\rho(\lambda | a,b) \neq \rho(\lambda).
\label{eq:no-hvi}
\end{align}
The formal expression in Eq.~\eqref{eq:no-hvi} is also known as superdeterminism. There are various proposals regarding the physical implications of it. Specifically, one might want to argue in favor of it on the grounds of the PBR theorem.

We here refrain from delving into more details. In particular, we do not intend to reflect on possible interpretations within the context of usual Bell type experiments. However, we shall note that arguments have been provided that a violation as in Eq.~\eqref{eq:no-hvi} may be relevant in quantum gravity \cite{tHooft:2001vwq,tHooft:2014znz}. 

The proposed picture in the present article, indeed, differs in many aspects. Importantly, in our understanding, it will be highly necessary to decouple from a physical (i.e.~gravitational) spacetime based picture, at least to a certain extent. An appropriate setting to examine is the quantum black hole in the light of holography, as discussed in Sec.~\ref{sec:mot} and detailed out further in Sec.~\ref{sec:inter}. This is what we are going to work out in Secs.~\ref{sec:tangle-hvs} and \ref{sec:eqbh}.

\section{Entanglement and hidden variables}
\label{sec:tangle-hvs}

The exterior of a quantum black hole may be divided into the following regions
\begin{align}
\mathbf s:\  r_\text{S} < r  < r_\text{S} + \ell_{s}, \qquad
\mathbf a:\ r_\text{S} + \ell_{s} < r < 3 r_\text{S} /2, \qquad
\mathbf f:\ r  > 3 r_\text{S}.
\label{eq:divs}
\end{align}
The Schwarzschild radius $r_\text{S} = 2 G M$ is proportional to the black hole mass $M$. The radial extension of the stretched horizon $\mathbf s$ determined by $\ell_{s}$ will be specified later. The region $\mathbf a$ defined in Eq.~\eqref{eq:divs} is called black hole atmosphere or zone. From now on, $\ell$ shall denote the distance from the black hole horizon located at $r=r_\text{S}$. The union $\mathbf{s a}$ shall be referred to as the quantum atmosphere that may be defined through $\epsilon < \ell$, where we have in mind a cutoff parameter $\epsilon \ll \ell_s$. More details regarding $\epsilon$ are elaborated further below and in Sec.~\ref{sec:eqbh}.

For the Schwarzschild black hole, the gravitational potential peaks at the boundary of region $\mathbf{sa}$, i.e.~$r = 3 G M$, which is known as the photon sphere. Quantum field theoretic radiation modes may accumulate across $\mathbf a$ before being excited and escaping into the far exterior region $\mathbf f$ in form Hawking quanta.

To leading order, the entropy of a black hole scales with the area of its horizon. The holographic principle argues that the horizon processes related information. Complementarity, as discussed in Sec.~\ref{sec:mot}, asserts that information associated with the initial pure state gets reflected in the near horizon region. 

How quantum information makes it to the vicinity of the horizon, and how this leads to the intricate entanglement between the black hole system and radiation must be understood. Be noted that associating a certain system with the near horizon dynamics and working out various quantum aspects thereof cannot unveil the underlying mechanism.\footnote{The model quantum system may even consist of extended objects such as strings and branes. 
Some related aspects shall be discussed elsewhere \cite{Akal:prep}.
}

The aim here is not to describe the dynamics of the evaporation process or to inspect the difficulty of decoding the information encoded in the outgoing radiation quanta. The seeming hardness of information decoding may, for instance, be linked to the enormous complexity of relevant operations \cite{Hayden:2007cs,Harlow:2013tf,Susskind:2013aaa}.

In order to eliminate unwanted complications, we shall focus on the black hole's initial phase. Consider a single quantum mode associated with $\mathbf a$ that can be excited as a radiation particle, which, in the following, shall be labeled by some spin operator $\sigma$. In the case of photons, we shall think about polarization instead of spin. 

Nonetheless, the labeling details are not important for the present discussion. We may generally think in terms of qubits. This shall even be so, when we refer to quantum modes associated with the stretched horizon $\mathbf s$. After being excited, $\sigma$ may escape into $\mathbf f$. In the semiclassical, global picture, $\sigma$ is viewed to be entangled with its partner particle in the interior. 

Demanding unitarity in a way consistent with the holographic principle, however, requires information processing to be traced back to the microscopic vicinity of the horizon. In the scenario described above, we may initially imagine some qubit assigned to $\mathbf s$, which is quantum correlated with the radiation mode in $\mathbf a$. The boundary of $\mathbf s$, operating as a quantum membrane, cf. e.g. \cite{tHooft:1984kcu,Akal:qes}, would therefore bear the underlying microstructure processed by the horizon. We may consider a total number of unscrambled black hole qubits of order $K = S_\text{BH}$ assigned to $\mathbf s$. 

The main statement is that such near horizon quantum mechanics is emergent in the sense that associated correlations can be explained in terms of hidden variables $\lambda_h$ assigned to the horizon.

Remember that in the standard Bell type scenario, the implications of Eq.~\eqref{eq:no-hvi} violating hidden variable independence in Eq.~\eqref{eq:hvi} can be interpreted in two directions. Either the state of the qubit, that is measured by some experimenter, decides about the setting (basis) of the measurement apparatus, or the other way around. Both would correlate with the hypothetical hidden variables $\lambda$ that clearly do not need to be assigned to a specific region in spacetime, especially, not at the spot where the respective EPR pair was created.

Coming back to the present situation, the entanglement between the qubit in $\mathbf s$ and the mode in $\mathbf a$ is argued to be compatible with a distribution function $\rho$ satisfying
\begin{equation}
\rho(\lambda_h | a_s ,b_a) = \rho(\lambda_h | a_s)  \neq \rho(\lambda_h).
\label{eq:e-hvi}
\end{equation}
According to the proposal in Eq.~\eqref{eq:e-hvi}, which is dubbed emergent hidden variable independence, the distribution of $\lambda_h$ depends on the setting $a_s$ or vice versa. The parameter $a_s$ may describe the emergent quantum properties of the stretched horizon $\mathbf s$, which would be dictated by the variables $\lambda_h$. Formally, we can therefore write 
\begin{equation}
a_s \neq a_s(\lambda_h) \Leftrightarrow
\lambda_h \neq \lambda_h(a_s).
\label{eq:a-lambda}
\end{equation} 

Even though fundamentally different, the expression in Eq.~\eqref{eq:a-lambda} shares certain parallels with Eq.~\eqref{eq:hvi} implying the mutual dependence between apparatus and EPR qubit, as both are hypothesized to be correlated with $\lambda$. However, recall that the pair of qubits in the latter case would have been created from a singlet state and viewed to be measured by apparatuses set up by two experimenters situated in arbitrarily different spacetime locations, both being correlated with the same variables $\lambda$.

We have to turn away from this picture. In the present situation, one of the \textit{apparatuses} in question, call it $D_s$, shall be identified as the region defined through $\ell < \ell_s$. The emergent black hole qubit, eventually, spreads over distances $\epsilon <\ell< \ell_s$. Considered to be the quantum mode of $D_s$, upon having been \textit{measured} by $D_s$, it would be correlated with the proposed variables $\lambda_h$.

On the other hand, Eq.~\eqref{eq:e-hvi} also implies that the setting $b_a$, say of an apparatus $D_A$ utilized to measure the black hole qubit just mentioned, or, equivalently, entangle with the mode of apparatus $D_s$ (i.e.~region $\mathbf s$) by some agent $A$ having control over region $\mathbf f$, is not correlated with $\lambda_h$. Formally, this can be expressed as follows
\begin{equation}
b_a = b_a(\lambda_h) \Leftrightarrow
\lambda_h = \lambda_h(b_a).
\label{eq:b-lambda}
\end{equation} 
One may consider $\mathbf a$, having its mode, i.e.~the $\mathbf a$-qubit we were referring to so far, being maximally entangled with the said black hole qubit assigned to $\mathbf s$ upon the described \textit{measurement process}, to be the apparatus $D_A$ of agent $A$. We may even say that $D_A$ is $A$.
In other words, Eq.~\eqref{eq:e-hvi} would respect a standard quantum theoretic formulation in the joint exterior region $\mathbf{a f}$, where the state of $\mathbf a$ will be characterized by the setting parameter $b_a$ obeying Eq.~\eqref{eq:b-lambda}.

For instance, one may consult a third, less powerful agent, call it the physical experimenter, who may live in $\mathbf f$ and might set up a conventional device to perform a quantum measurement upon the escaping radiation particle that has been excited in $\mathbf a$.

Note that, initially, we assume that all $K$ mode entanglements between $\mathbf s$ and $\mathbf a$ are pairwise maximal. This would be compatible with a local quantum field theory defined in the exterior. Importantly, it shall be emphasized that the setting $a_s$ satisfying Eq.~\eqref{eq:a-lambda} cannot be accessed by an agent. This is far different from the usual scenario in a Bell type experiment. The range of the validity of violating Eq.~\eqref{eq:hvi} in the form of Eq.~\eqref{eq:e-hvi} may be restricted to distances 
\begin{equation}
\ell < \epsilon,
\label{eq:ell-eps}
\end{equation}
where $\epsilon \ll \ell_s$. More details regarding the cutoff parameter $\epsilon$ appearing in Eq.~\eqref{eq:ell-eps} are discussed later. For instance, in the string theory context, the parameter $\ell_s$ may be of order the string length, see Sec.~\ref{sec:eqbh}. Note that $\ell_s > \ell_\text{P}$.

We have pointed out that information holographically processed by the black hole horizon would be projected onto emergent qubits having support in region $\mathbf s$, which would be correlated with hidden variables $\lambda_h$ operating across sub-Planckian distances from the horizon, i.e.~$\ell < \epsilon$. This would give rise to quantum entanglement between the black hole system, i.e.~qubits in  $\mathbf s$, and field theoretic radiation modes in $\mathbf a$ that can be excited and escape into the far exterior region $\mathbf f$. 

The condition of emergent hidden variable independence, Eq.~\eqref{eq:e-hvi}, respects standard quantum theory in the exterior. The correlations between $\lambda_h$ and $\mathbf s$, however, turn out to be necessary due to the holographic processing of information by the horizon. A nonprobabilistic, hidden variable dynamics associated with the black hole horizon may thus lead to the desired violation of the Bell inequalities and explain quantum mechanical expectation values consistent with the exterior description.

From the distance, the boundary of $\mathbf s$ may be viewed as a quantum membrane \cite{Akal:qes} so that the physical Hilbert space at a given time may factorize as $\mathcal{H}_\text{exterior} = \mathcal{H}_\mathbf{s} \otimes \mathcal{H}_\mathbf{a} \otimes \mathcal{H}_\mathbf{f}$. Taking $|\mathcal{H}|$ to be the dimensionality of a given Hilbert space $\mathcal H$, $\log |\mathcal{H}_\mathbf{s}|$ and $\log |\mathcal{H}_\mathbf{a}|$ would initially be proportional to $S_\text{BH}$.
During evaporation, one may then distinguish between the two phases of the quantum black hole. Namely, the latter is viewed to be young if $|\mathcal{H}_\mathbf{f}| < |\mathcal{H}_\mathbf{s}| |\mathcal{H}_\mathbf{a}|$. Oppositely, when the black hole is referred to be old, i.e.~passing over the Page time \cite{Page:1993df, Page:1993wv}, one would have $|\mathcal{H}_\mathbf{f}| > |\mathcal{H}_\mathbf{s}| |\mathcal{H}_\mathbf{a}|$.

In summary, we have argued that the black hole horizon may be crucial in restoring an ontology in quantum mechanics, that is compatible with the idea of spacetime emergence as envisaged by the holographic principle. However, we have not yet presented a specific argument for why we expect the proposed hidden variables $\lambda_h$ to operate on the black hole horizon or even across sub-Planckian distances from it. This is the content of Sec.~\ref{sec:eqbh}. Before doing so, we would like to elaborate on a recent no-go theorem in Sec.~\ref{sec:exwigner}, that we will be referring to later.

\section{Wigner, his twin, and their friends}
\label{sec:exwigner}

Consider $\mathfrak{T}$ to be a theory having the following three properties. 

($\mathsf{Q}$) $\mathfrak{T}$ respects the standard rules of quantum mechanics as sketched in Sec.~\ref{sec:qm}. For instance, agent $A$ is certain that a system is in an eigenstate of some observable $O$ with eigenvalue $\Theta$ at time $t_0$. When a measurement of $O$ is done at time $t_1 > t_0$, $A$ is certain that the measurement outcome $a$ is $\Theta$ at time $t_1$. By certainty it is meant that the probability $p$ assigned to a given proposition by the Born rule is $p=1$.

($\mathsf{C}$) $\mathfrak{T}$ requires logical consistency for statements about measurement outcomes. These may even hold when perspectives of different agents are considered. For instance, if agent $A$ is certain that at time $t$ another agent $A'$ is certain that $a=\Theta$ in the sense of ($\mathsf{Q}$), then $A$ is certain that $a=\Theta$ at time $t$.

($\mathsf{S}$) $\mathfrak{T}$ rules out more than one single outcome if an agent performs a measurement. For instance, if agent $A$ is certain that $a=\Theta$ is true, then $A$ is certain that $a \neq \Theta$ is not true.

Assume the existence of four agents: Wigner ($W$), Wigner's twin ($\overline W$), the friend of $W$ ($F$), and the friend of $\overline W$ ($\overline F$). The setup is the following. Agents $\overline F$ and $F$ both perform measurements on their two level quantum systems; a quantum coin $C$ with orthonormal basis $| \textsf{h} \rangle$ and $|\textsf{t} \rangle$ and a particle $S$ with eigenstates $| \uparrow \rangle$ and $| \downarrow \rangle$, respectively.
Agents $\overline W$ and $W$ perform measurements on the combined systems $C \overline F$ and $S F$, respectively. Before starting the experiment, $C$ is prepared in the state $\sqrt{\frac{1}{3}} | \mathsf{h} \rangle_C + \sqrt{\frac{2}{3}}  | \mathsf{t} \rangle_C$ modeling a random coin flip. 
It is assumed that $t_0 < t_1 < t_2 < t_3 < t_4$. The experiment proceeds as follows.

At time $\pmb{t=t_0}$, agent $\overline F$ measures $C$ and records the result $\overline f = \mathsf{heads}$ or $\overline f =\mathsf{tails}$, being stored in a memory state $| \overline f \rangle_{\overline F}$.

At time $\pmb{t=t_1}$, agent $\overline F$ prepares the state of particle $S$ according to the result $\overline f$ at time $t_0$. If $\overline f =\mathsf{heads}$, then $S$ is prepared in state $| \downarrow \rangle_S$. If $\overline f = \mathsf{tails}$, then $S$ is prepared in state $| \rightarrow \rangle_S \equiv \sqrt{\frac{1}{2}} ( | \uparrow \rangle_S + | \downarrow \rangle_S )$. Afterwards, agent $\overline F$ sends $S$ to agent $F$.

At time $\pmb{t=t_2}$, agent $F$ measures the spin $z/2$ (here, $\hbar = 1$ is set) of particle $S$, i.e. $z = \pm$, in the basis $\{ | \uparrow \rangle, | \downarrow \rangle  \}$ and records the result, stored in a memory state $| z \rangle_F$.

At time $\pmb{t=t_3}$, agent $\overline W$ measures the combined system $C \overline F$ in the basis $\{ | \overline{\mathsf{ok}} \rangle_{\overline F C} ,  | \overline{\mathsf{fail}} \rangle_{\overline F C} \}$, where $| \overline{\mathsf{ok}} \rangle_{\overline F C} \equiv \sqrt{\frac{1}{2}} (|\mathsf{heads}\rangle_{\overline F} |\textsf{h}\rangle_C - |\mathsf{tails}\rangle_{\overline F} |\textsf{t}\rangle_C)$ and $| \overline{\mathsf{fail}} \rangle_{\overline F C} \equiv \sqrt{\frac{1}{2}} (|\mathsf{heads}\rangle_{\overline F} |\textsf{h}\rangle_C + |\mathsf{tails}\rangle_{\overline F} |\textsf{t}\rangle_C)$. Depending on the outcome, the result $\overline w$ is recorded as either $\overline w = \overline{\mathsf{ok}}$ or $\overline w = \overline{\mathsf{fail}}$.

At time $\pmb{t=t_4}$, agent $W$ measures the combined system $S F$ in the basis $\{ | \mathsf{ok} \rangle_{F S} ,  | \mathsf{fail} \rangle_{F S} \}$, where $| \mathsf{ok} \rangle_{F S} \equiv \sqrt{\frac{1}{2}} (|-\rangle_{F} |\downarrow\rangle_S - |+\rangle_{F} |\uparrow\rangle_S)$ and $| \mathsf{fail} \rangle_{F S} \equiv \sqrt{\frac{1}{2}} (|-\rangle_{F} |\downarrow\rangle_S + |+\rangle_{F} |\uparrow\rangle_S)$. Depending on the outcome, the result $w$ is recorded as either $w = \mathsf{ok}$ or $w = \mathsf{fail}$.
After all steps are completed, $W$ and $\overline W$ compare their results $w$ and $\overline w$, respectively. The experiment repeats so long until it stops when $w = \mathsf{ok}$ and $\overline w = \overline{\mathsf{ok}}$. 

Now, if such an experiment is described by theory $\mathfrak{T}$, it turns out that this leads to contradictory statements; the Frauchiger--Renner (FR) no-go theorem states that no such theory can have all three properties $\mathsf{Q}$, $\mathsf{C}$, and $\mathsf{S}$ simultaneously \cite{frauchiger2018quantum}. According to the authors, various interpretations of quantum theory seem not to be in agreement on the origin of the contradiction. They necessarily conflict at least with one of the properties above. Refer, for instance, to \cite{sudbery2019hidden} for further inspections along this direction. 

We here do not intend to delve into the interpretational aspects. The goal of the present work is not to advocate an interpretational resolution to the problem we will be focusing on. Whenever we refer to standard quantum theory, we suffice with accepting the Copenhagen interpretation. 

Taking this view, the self referential use of $\mathfrak{T}$ in the experiment above tells us that either $S$, the combined system (i.e.~laboratory) $LAB = S F$ ($\otimes$ measurement apparatus $D_F$), or the theory, which agent $W$ has control over, cannot be described within the standard Copenhagen formulation. The same applies for the side of agent $\overline W$. Before transitioning to the next part, we shall note that modifications rather than interpretations, like certain hidden variable models, even though being criticizable on other grounds, may invalidate the FR theorem \cite{Sudbery2017SingleWorldTO}. 

\section{Eternal quantum AdS black hole revisited}
\label{sec:eqbh}

As in the previous scenario, we similarly reason below that contradictory statements can be inferred in the case of the two sided quantum black hole discussed in Sec.~\ref{sec:inter}.  Especially, we provide additional arguments for why the hidden variables $\lambda_h$, proposed to be accountable for the emergence of the near horizon quantum mechanics in Sec.~\ref{sec:tangle-hvs}, are operative on the black hole horizon. Based on this, we additionally elaborate on the possibility of how the notion of computational complexity may play a role in censoring the implied sub-Planckian actuality from physical observers in a gravitational spacetime. 

\subsection{Setting up the basics}

Consider the TFD state defined on the boundaries of the eternal black hole. The microstates may be modeled by a total number of qubits of order $K = S_\text{BH}$. At time $t=0$, the bulk system may consist of a collection of unscrambled qubits making up the stretched horizons $\mathbf s_{L}$ and $\mathbf s_{R}$. By construction, every such black hole qubit will initially be entangled with a partner qubit assigned to the stretched horizon on the opposite side. In other words, we assume the absence of vertical correlations, cf.~e.g.~\cite{Susskind:2013lpa,Susskind:2013aaa}. 

For simplifying reasons, we shall focus on a single pair of entangled qubits associated with $\mathbf s_{L}$ and $\mathbf s_{R}$. Nonetheless, the total state describing all pairwise maximally entangled qubits may be written as $| \text{Bell} \rangle_s^{\otimes K}$. The overall time evolution may be modeled by a quantum circuit, where the gates, chosen from a universal set, may (randomly) act between pairs of qubits on each side, sequentially or in parallel, cf.~\cite{Hayden:2007cs}.

The theory describing the two laboratory systems, which are denoted in the following by $LAB_L$ and $LAB_R$, is given by the boundary theory. On left side, CFT$_L$ may describe the combined system consisting of a black hole qubit in its possible states $\{ | \overline 0 \rangle_{s_L},  | \overline 1 \rangle_{s_L} \}$, and the apparatus (or memory) identified as $\mathbf a_L$ with its possible states $\{ |\overline{\mathsf{null}} \rangle_{a_L},  | \overline{\mathsf{one}} \rangle_{a_L} \}$ to which some agent $\overline F$, having control over $\mathbf f_L$, may have access to. 

We shall not distinguish between agent and memory, as in Sec.~\ref{sec:tangle-hvs}. Whenever agent $\overline F$ measures a black hole qubit, i.e.~entangles a mode in $\mathbf a_L$ with a qubit in $\mathbf s_L$, it is assumed that the atmosphere modes become pairwise maximally entangled with the qubits assigned to the stretched horizon. This shall apply to an ample amount of qubits in a given, sufficiently short time interval. Agent $\overline F$ is assumed to be skilled enough to distinguish between those qubit entanglements.  

The same inferences also apply in the case of CFT$_R$ describing the right black hole qubit being in the possible states $\{ | 0 \rangle_{s_R}, | 1 \rangle_{s_R} \}$, as well as the mode in $\mathbf a_R$ with its possible states $\{ | \mathsf{null} \rangle_{a_R}, | \mathsf{one} \rangle_{a_R} \}$. Together they shall define the laboratory system $LAB_R$ assigned to agent $F$, who is assumed to be equally skilled and has access to the own measurement apparatus identified as $\mathbf a_R$.

The dependence of the laboratory states on the black hole qubit states can be written in form of linear maps of the form
\begin{align}
\begin{split}
| \overline 0 \rangle_{s_L} 
&\mapsto 
| \overline{\mathsf{ok}} \rangle_{LAB_L}
\equiv 
| \overline 0 \rangle_{s_L}
\otimes 
| \overline{\mathsf{null}} \rangle_{a_L},\\
| \overline 1 \rangle_{s_L} 
&\mapsto 
| \overline{\mathsf{fail}} \rangle_{LAB_L}
\equiv 
| \overline 1 \rangle_{s_L}
\otimes 
| \overline{\mathsf{one}} \rangle_{a_L},\\
| 0 \rangle_{s_R} 
&\mapsto 
| \mathsf{fail} \rangle_{LAB_R}
\equiv 
| 0 \rangle_{s_R}
\otimes 
| \mathsf{null} \rangle_{a_R},\\
| 1 \rangle_{s_R} 
&\mapsto 
| \mathsf{ok} \rangle_{LAB_R}
\equiv 
| 1 \rangle_{s_R}
\otimes 
| \mathsf{one} \rangle_{a_R},
\end{split}
\label{eq:linmaps}
\end{align}
where $LAB_L \equiv s_L a_L$ and $LAB_R \equiv s_R a_R$, respectively.

Consider a time window $t_0 \leq t \leq t_3$, where $t_3 - t_0$ shall be such that at $t= t_3$ the specified entangled qubits by agents $\overline F$ and $F$ may still be viewed unscrambled. One may imagine that at $t=t_3$, the qubits under consideration have still not been acted on by gates in a (sequential) quantum circuit. Recall that we start with the purely horizontally correlated state at $t=0$, thus we set $t_0=0$. The relevant time steps are organized such that $0  < t_1 < t_2 < t_3$.

\subsection{An experiment}

In the following, we demonstrate that the setup outlined above yields contradictory statements, if the properties $\mathsf{Q}$, $\mathsf{C}$, and $\mathsf{S}$ are assumed. Consider the initial state assigned to the left qubit $\left(  |\overline{1} \rangle_{s_L} + | \overline{0} \rangle_{s_L} \right)/\sqrt{2}$. Since, at this stage, the latter is maximally entangled with the partner qubit on the right side, we have a Bell state of the form $| \text{Bell} \rangle_s \equiv \left( | \overline{0} \rangle_{s_L} | 1 \rangle_{s_R} +|\overline{1} \rangle_{s_L} | 0 \rangle_{s_R} \right)/\sqrt{2}$ at the level of the bulk agents $\overline F$ and $F$. 

We remark that the convention used here can be changed so that $| \overline 0 \rangle_{s_L} \rightarrow | 1 \rangle_{s_L}$ and $| \overline 1 \rangle_{s_L} \rightarrow | 0 \rangle_{s_L}$ which will have no impact on what we are going to infer. The states of the bases for conducting measurements on the laboratory systems by boundary agents $\overline W$ and $W$ are specified as $\{ | \overline{\mathsf{fail}} \rangle_{LAB_L}, | \overline{\mathsf{ok}} \rangle_{LAB_L}  \}$ and $\{ | \mathsf{fail} \rangle_{LAB_R} , | \mathsf{ok} \rangle_{LAB_R}  \}$, respectively, see Eq.~\eqref{eq:linmaps}.

Now, suppose that the measurement outcome $\overline{\mathsf{one}}$ occurs at time $t=t_0$, which shall be stored in the memory state $| \overline{\mathsf{one}} \rangle_{a_L}$ by agent $\overline F$.
Then, due to $\mathsf{Q}$, $F$ must conclude the result $\mathsf{null}$ at time $t=t_1$, hence
\begin{equation}
\overline{\mathsf{one}}\ \text{at}\ \pmb{t=t_0} \Rightarrow\ \mathsf{null}\ \text{at}\ \pmb{t=t_1}.
\label{eq:imply_0}
\end{equation}
Agent $F$ stores the outcome at $t=t_1$ given in Eq.~\eqref{eq:imply_0} in the memory state $| \mathsf{null} \rangle_{a_R}$. 

On the other hand, the state describing the combined system $s_R LAB_L$, where $LAB_L$ captures the state of the black hole qubit and the memory state on the left side, is
\begin{equation}
\sqrt{\frac{1}{2}}|\overline 0 \rangle_{s_L} |\overline{\mathsf{null}} \rangle_{a_L} | 1\rangle_{s_R} 
+ 
\sqrt{\frac{1}{2}} | \overline{\mathsf{fail}} \rangle_{s_L a_L} | 0\rangle_{s_R},
\label{eq:GHZ}
\end{equation}
which basically (after changing the convention as mentioned) corresponds to a Greenberger--Horne--Zeilinger (GHZ) state. Hence, if $\mathsf{null}$ occurs at time $t=t_1$ upon $F$'s measurement, the outcome of measuring the system $LAB_L$ at $t=t_2$ by $\overline W$ must be $\overline{\mathsf{fail}}$ according to Eq.~\eqref{eq:GHZ}, i.e.
\begin{equation}
\mathsf{null}\ \text{at}\ \pmb{t=t_1} \Rightarrow\ \overline{\mathsf{fail}}\ \text{at}\ \pmb{t=t_2}.
\label{eq:imply_1}
\end{equation}
$\overline W$ stores the outcome in Eq.~\eqref{eq:imply_1} at time $t=t_2$ in a memory state $| \overline{\mathsf{fail}} \rangle_{LAB_L}$.

When a measurement of $s_R$ by $F$ takes place, means $a_R$ becomes entangled with $s_R$, the state of the combined system $LAB_L LAB_R$ takes the general form
\begin{equation}
\sqrt{\frac{1}{2}}|\overline 0 \rangle_{s_L} |\overline{\mathsf{null}} \rangle_{a_L} | 1\rangle_{s_R}  | \mathsf{one} \rangle_{a_R}
+ 
\sqrt{\frac{1}{2}} | \overline{\mathsf{fail}} \rangle_{LAB_L} | 0\rangle_{s_R} | \mathsf{null} \rangle_{a_R}.
\label{eq:4-part}
\end{equation}
Since we take into account local entanglement with both atmosphere regions, this differs from the three partite GHZ state discussed, for instance, in \cite{Susskind2014AddendumTC}. The state in Eq.~\eqref{eq:4-part}, re-expressed in terms of the measurement bases states, reads
\begin{align}
| \text{Bell} \rangle_{LAB} \equiv \sqrt{\frac{1}{2}} \left( | \overline{\mathsf{ok}} \rangle_{LAB_L} | \mathsf{ok} \rangle_{LAB_R} 
+ | \overline{\mathsf{fail}} \rangle_{LAB_L} | \mathsf{fail} \rangle_{LAB_R}\right).
\label{eq:Bell-lab}
\end{align}
Eq.~\eqref{eq:Bell-lab} implies the possibility
\begin{equation}
\overline{\mathsf{ok}}\ \text{at}\ \pmb{t=t_3}\ \wedge\ \mathsf{ok}\ \text{at}\ \pmb{t=t_3}
\label{eq:poss_1}
\end{equation} 
upon the measurements of boundary agents $\overline W$ and $W$, respectively. 

Notably, if a measurement is performed on system $LAB_R$ by $W$ at time $t=t_3$, it may not affect the state of system $LAB_L$. Due to $\mathsf{S}$, one may, therefore, expect the outcomes to be the same, if a measurement is performed on system $LAB_L$ by $\overline W$ at times $t=t_2 $ and $t=t_3$. Taking into account Eq.~\eqref{eq:poss_1} thus implies the possibility that
\begin{equation}
\overline{\mathsf{ok}}\ \text{at}\ \pmb{t=t_2}
\label{eq:poss_2}
\end{equation}
is stored in $\overline W$'s memory state $| \overline{\mathsf{ok}} \rangle_{LAB_L}$.

However, given that $\mathsf{C}$ holds, Eq.~\eqref{eq:poss_2} would contradict the implication in Eq.~\eqref{eq:imply_1}, according to which the memory accessed by $\overline W$ stores the state $| \overline{\mathsf{fail}} \rangle_{LAB_L}$. We thus encounter that assuming $\mathsf{Q}$, $\mathsf{C}$, and $\mathsf{S}$ does result in contradictory statements between the boundary and bulk descriptions.

\subsection{Nonquantumness}
\label{subsec:nonqu}

What do we infer from the contradiction above? It is clear that invalidating it can be achieved by abandoning at least one of the assumptions $\mathsf{Q}$, $\mathsf{C}$, or $\mathsf{S}$. However, the two sided system has a dual quantum mechanical boundary description by construction, which is associated with the laboratory systems $LAB_{L}$ and $LAB_{R}$. In contrast to the extended Wigner's friend scenario, this property is not left up to one's choice. Respecting all three assumptions will not cause any tension if considerations are restricted to the boundary description. 

Nevertheless, we argue below that standard quantum theory cannot be applicable universally in the bulk. More specifically, a breakdown is reasoned with respect to the modes assigned to $\mathbf s_{L}$ and $\mathbf s_{R}$. First of all, such an assertion may be traced back to the arguments raised in Sec.~\ref{sec:tangle-hvs}. Recall that according to the proposal of emergent hidden variable independence, Eq.~\eqref{eq:e-hvi}, the modes assigned to $\mathbf a$ may have a standard description, whereas the quantum mechanics of $\mathbf s$ is argued to be emergent. 

Notably, operators associated with $\mathbf a$ can relatively easily be reconstructed from the boundary CFT, since they turn out to require a moderate level of complexity, as will be detailed out shortly. However, the same computational easiness cannot be inferred in the case of $\mathbf s$ having thickness $\ell_s$. This, particularly, appears to be so when sub-Planckian distances, i.e.~$\ell \ll \ell_\text{P}$, are in question. Note that $\ell_\text{P} < \ell_s$ in general.

Remember that the hidden variables $\lambda_h$ associated with the emergent quantum mechanics are assigned to the black hole horizon. Since the latter lies to the infinite future of the CFT, $t=\infty$, those variables cannot be assigned to a finite time interval on the AdS boundary. The relationship between the near horizon region and large boundary times has, for instance, been discussed in \cite{Susskind:2013lpa}. 

Nevertheless, we here do not require the relevance of the interior for holding up unitarity. A more adequate picture in the present context would be saying that the complexity associated with the sub-Planckian vicinity of the horizon is such, that it would (effectively) require smearing over an infinite time interval on the AdS boundary to reconstruct associated information or an operator of equal complexity. 

Accordingly, because of such constraints imposed by the time dimension $t$, it would not be possible to implement certain hidden variables $\lambda_\text{CFT}$ assigned to the asymptotic boundary, which, otherwise, might be one-to-one mapped to the variables $\lambda_h$. Formally, we can re-express this statement as
\begin{equation}
\mathcal{H}_\text{CFT} \neq \mathcal{H}_\text{CFT}(\lambda_\text{CFT}),
\label{eq:no-cft-hvs}
\end{equation}
alluding to the nonexistence of a hidden variable formulation on the boundary.

On the other hand, one might want to implement a set of bulk hidden variables $\lambda_\mathbf{af}$ to explain quantum correlations across the joint bulk exterior region $\mathbf a \mathbf f$. However, be noted that boundary reconstructions of bulk operators located across $\mathbf a \mathbf f$ are moderately complex, which, in fact, would contradict the claim in Eq.~\eqref{eq:no-cft-hvs}, since a direct map between some hypothetical variables $\lambda_\text{CFT}$ and $\lambda_\mathbf{a f}$ may then exist. 

It is of particular significance emphasizing that in the gravitational bulk spacetime no hidden variables are required. Enforcing them anyway would necessarily result in a nonlocal formulation, which is not what one wants to aim for. 

In other words, once a smooth physical spacetime has emerged, there will be no way and, in fact, no necessity to provide a formulation by means of hidden variables, which preserves the desired notion of locality, hence
\begin{equation}
\mathcal{H}_\mathbf{af} \neq \mathcal{H}_\mathbf{af}(\lambda_\mathbf{af}).
\label{eq:no-af-hvs}
\end{equation}
A description in terms of standard quantum theory may suffice in such instances. These arguments would equally apply to the asymptotic lower dimensional boundary spacetime, which may also reinforce the statement in Eq.~\eqref{eq:no-cft-hvs}. 

Nonetheless, it is clear that from the exterior point of view, the black hole horizon is different in the sense that it processes information associated with a system that does not have a spacetime based description. We formally express these statements as
\begin{equation}
\mathcal{H}_\text{exterior} = \mathcal{H}_\mathbf{s}(\lambda_h) \otimes \mathcal{H}_\mathbf{a f}.
\label{eq:s-af}
\end{equation}
The emergent near horizon Hilbert space structure may thus allow for restoring the required complete operator algebra discoursed in Sec.~\ref{sec:inter}. 

\subsection{Planckian projections, uncomputability, and infinite future}

We shall first introduce some relevant equations related to operator complexities associated with the bulk-boundary map in AdS/CFT. Consider $\ell$ to be again the distance from the black hole horizon and assume that $\ell_s \gg \ell$. A general relation of the form $\ell = \ell_s \exp(- \omega) $ can be derived, where the coordinate variable $\omega = \frac{t}{L_\text{AdS}}$ denotes the dimensionless Rindler time, $L_\text{AdS}$ stands for the AdS curvature radius, and $t$ is the usual (AdS) Schwarzschild time. The boundary of the stretched horizon $\ell_s$ may lie at a distance $L_\text{AdS}$ from the actual horizon at $r=r_\text{S}$, so that $\ell = L_\text{AdS} \exp(-\omega)$. 

In order to make the setup more explicit, we may focus on a unit black hole. In other words, we consider the limit where the black hole transitions to a single long string of length $K \ell_\text{string}$ \cite{Susskind:2013lpa}, with the 't Hooft coupling $g^2 N$ being of order unity \cite{tHooft:1973alw}. Note that the parameter $\ell_\text{string}$ corresponds to the length of the fundamental string, i.e.~$\ell_\text{string} = \sqrt{\alpha'}$. Viewing the system under consideration to be slightly above the Hawking--Page transition, we may further set $r_\text{S} \simeq \ell_s\simeq \ell_\text{string}$. Recall that $S_\text{BH} \propto r_\text{S}^2/ \ell_\text{P}^2 $. 

At a Planckian distance from the horizon, $\ell = \ell_\text{P}$, one arrives at the scrambling time $t_\text{scr} = L_\text{AdS} \omega_\text{scr} = L_\text{AdS}  \log(S_\text{BH})$ \cite{Sekino:2008he}. Taking into account general features of (parallel) quantum circuit growth, a relation between $\ell$ and quantum computational complexity $\mathcal{C}$ can be obtained \cite{Susskind:2013aaa,Susskind:2014rva},
\begin{equation}
\ell^2/ L_\text{AdS}^2 = \exp\left( -\mathcal{C} \right).
\label{eq:l-C-S}
\end{equation} 
We here ignore the $S_\text{BH}$ dependence of the exponent. The entropy $S_\text{BH}$, however, will reappear when specifying the length scales on the left hand side in Eq.~\eqref{eq:l-C-S}.

For instance, consider the case $\ell = \ell_*$, where
\begin{equation}
\ell_* \equiv L_\text{AdS} \exp\left( -e^{S_\text{BH}} \right).
\label{eq:ells}
\end{equation}
Remember that $\ell_s \sim \ell_\text{P} \sqrt{S_\text{BH}}$. At sub-Planckian distances of order $\ell_*$, Eq.~\eqref{eq:ells}, the formula in Eq.~\eqref{eq:l-C-S} yields $\mathcal{C} = \mathcal{C}_\text{max} \equiv \exp \left( S_\text{BH} \right)$, supported over all $K = S_\text{BH}$ qubits making up the black hole system. The value $\mathcal{C}_\text{max}$ is referred to as the maximum in the sense that any given unitary operation can be implemented by a quantum circuit consisting of no more than an exponential number of gates, i.e.~$\exp(K)$. Such an operation would be achieved by a time of order the recurrence time $t_\text{rec} = L_\text{AdS} \exp \left( S_\text{BH} \right)$. 

At the scrambling time $t = t_\text{scr}$, all $K$ qubits would be in indirect contact with each other, so that, following a linear growth, complexity would take the value $\mathcal{C}_\text{scr} = \log(S_\text{BH})$. For boundary times $t > t_\text{rec}$, complexity eventually stops increasing. Expectedly, after such large times the notion of a global bulk spacetime may not persist. 

However, recall that we here adopt the exterior standpoint. As discussed in Sec.~\ref{sec:inter}, the spacetime being dual to the complete boundary theory is geodesically incomplete. The infinite future of the CFT, i.e.~$t=\infty$, is represented in the bulk by the black hole horizon. Hence, from the far distance, one is in principal allowed to consider even smaller distances
\begin{equation}
\ell < \ell_*,
\label{eq:ell-ells}
\end{equation}
which would render complexity exceeding $\mathcal{C}_\text{max}$ according to the formula in Eq.~\eqref{eq:l-C-S}. 

Suppose that one wants to reconstruct certain bulk dynamics from the boundary CFT. According to the relations above, any associated operator assigned to $\mathbf s$, having eventually support across the sub-Planckian vicinity of the actual horizon, would have no complexity counterpart in the boundary theory. A correspondence of such type would only be maintained up to distances $\ell = \ell_*$. Remember that the boundary of $\mathbf s$ is located at $\ell = L_\text{AdS}$, where $L_\text{AdS} \gg \ell_*$. 
 
What has been discussed up to here shall now be linked to the preceding discussion in Sec.~\ref{subsec:nonqu}. Namely, given that a maximum complexity does apply, we may deduce that reconstructing the dynamics of region $\mathbf s$ at distances $\ell < \ell_*$ in terms of boundary operations smeared over a finite time interval cannot be achievable computationally. Because of such a computational barrier, the proposed hidden variables $\lambda_h$ may be allowed to have imprints across sub-Planckian distances from the black hole horizon. 

In fact, as expressed by Eq.~\eqref{eq:no-cft-hvs}, if the notion of a maximum complexity would not be applicable, one would computationally be able to reconstruct operators having such close support from the horizon. Accordingly, a direct mapping between $\lambda_h$ and some hypothetical variables $\lambda_\text{CFT}$ assigned to the boundary may then exist. 

However, such a possibility has already been precluded by means of previous arguments, refer to Eqs.~\eqref{eq:no-cft-hvs} and \eqref{eq:no-af-hvs} resulting in the formal expression in Eq.~\eqref{eq:s-af}. Therefore, taking into account Eqs.~\eqref{eq:ell-eps} and \eqref{eq:ell-ells}, we can set $\epsilon = \ell_*$ for the value of the cutoff parameter introduced in Sec.~\ref{sec:tangle-hvs}. Hence, confining the computationally nonaccessible hidden variable dynamics to distances $\ell < \epsilon$ turns out to be reasonable from the complexity point of view either. 

We shall finalize the present discussion by pointing out a tempting possibility. As near horizon complexity can directly be related to time, this provides a basis for making connections with the arguments in Sec.~\ref{sec:tangle-hvs}. Recall that the proposal of emergent hidden variable independence, Eq.~\eqref{eq:e-hvi}, implies a kind of sub-Planckian determinism (in a logical sense). In light of this, we infer that uncomputability on quantum complexity grounds may eventually censor such sub-Planckian actuality from physical bulk observers perceiving the notion of a gravitational spacetime. Notably, in the particular case of AdS/CFT, the dual boundary operation of this censorship may correspond to projecting everything into the infinite future of the CFT.

\section{Conclusion}
\label{sec:conc}

Regardless of the countless interpretations and researchers' ability to devise a landscape of mathematical hocus pocus, nature does not appear to tolerate any sort of redundancy. 

Given that the holographic principle holds, quantum information eventually has to be lost inside black holes. However, the horizon holographically processing information would allow for the emergence of a quantum atmosphere maintaining unitarity, in reconciliation with the validity of local quantum field theory in the exterior. We have pointed out that such emergent near horizon quantum mechanics is compatible with a nonprobabilistic, hidden variable formulation, where the hidden variables are to be operative on the horizon of the black hole. We shall remark that a description via standard quantum theory may already break down across sub-Planckian distances from the latter. 

The present explorations suggest that spacetime horizons, in general, may play an essential role in restoring a long speculated ontology in quantum mechanics. Crucially, one must take into account that gravitational spacetime itself is an emergent and not a fundamental concept. Eventually, there may be neither gravity nor quantum in quantum gravity.

\section*{Acknowledgement}
I am grateful to
Nava Gaddam, Umut G{\"u}rsoy, and Tadashi Takayanagi 
for their comments on the draft.


\bibliographystyle{JHEP}
\bibliography{BQMGH}

\end{document}